\documentclass[conference, a4paper]{IEEEtran}
\usepackage{amsmath}
\usepackage{amssymb}  
\usepackage{mathtools} 
\usepackage{graphicx} 
\usepackage{subcaption} 
\usepackage{float} 
\usepackage{cite}
\usepackage{xcolor}
\usepackage[top=19mm, bottom=45mm, left=13mm, right=13mm]{geometry}
\title{Enhanced Open-Source NWDAF for Event-Driven Analytics in 5G Networks}

\author{
    \IEEEauthorblockN{Henok~Daniel\textsuperscript{\dag}, Omar~Alhussein\textsuperscript{\dag}, 
    Jie~Liang\textsuperscript{\ddag},
    Cheng~Li\textsuperscript{\ddag},
    and Ernesto~Damiani\textsuperscript{\P}\textsuperscript{\S}\\
    }
    \IEEEauthorblockA{\textsuperscript{\dag}KU 6G Research Centre, Department of Computer Science, Khalifa University, Abu Dhabi, UAE\\}
    \IEEEauthorblockA{\textsuperscript{\P}Center of Cyber-Physical Systems (C2PS), Department of Computer Science, Khalifa University, Abu Dhabi, UAE\\}
    \IEEEauthorblockA{\textsuperscript{\S}Department of Computer Science, Università degli Studi di Milano, Milano, Italy\\}
    \IEEEauthorblockA{\textsuperscript{\ddag}School of Engineering Science, Simon Fraser University, Vancouver, Canada\\
    Emails: \{henok.kahsay,~omar.alhussein,~ernesto.damiani\}@ku.ac.ae, \{jiel,~li\_cheng\}@sfu.ca\\}
}

\begin{document}
\maketitle
\footnotetext{ISBN 978-3-903176-72-0 © 2025 IFIP}

\begin{abstract}\label{sec-1}
The network data analytics function (NWDAF) has been introduced in the fifth-generation (5G) core standards to enable event-driven analytics and support intelligent network automation. However, existing implementations remain largely proprietary, and open-source alternatives lack comprehensive support for end-to-end event subscription and notification. In this paper, we present an open source NWDAF framework integrated into an existing Free5GC implementation, which serves as an open-source 5G core implementation. Our implementation extends the session management function to support standardized event exposure interfaces and introduces custom-built notification mechanisms into the SMF and the access and mobility management function for seamless data delivery. The NWDAF subscribes to events and generates analytics on user equipment (UE) behavior, session lifecycle, and handover dynamics.
We validate our system through a two-week deployment involving four virtual next-generation NodeBs (gNBs) and multiple virtual UEs with dynamic mobility patterns. To demonstrate predictive capabilities, we incorporate a mobility-aware module that achieves 80.65\% accuracy in forecasting the next gNB handover cell. The framework supports reliable UE registration, state tracking, and cross-cell handovers.
\end{abstract}

\section{Introduction}\label{sec-2}

Unlike previous generations, fifth-generation (5G) networks adopt a service-based architecture (SBA) that enables modular and scalable network functions, making it well-suited for a variety of applications, spanning enhanced mobile broadband, massive machine-type communications, and ultra-reliable low-latency communications \cite{34}. Towards data-driven network operations and intelligence, the network data analytics function (NWDAF) has been introduced in third generation partnership project (3GPP)-Release 15 \cite{14}. The NWDAF provides data-driven insights by collecting information from multiple network functions (NFs). Such capabilities support emerging technologies, such as autonomous systems, smart cities, and industrial automation \cite{35}.

To accelerate research and development in this space, open-source 5G projects have gained prominence, offering software-based implementations of core and radio access network (RAN) components. Projects such as Free5GC, Open5GS, UERANSIM, srsRAN, and OpenAirInterface (OAI) offer platforms for testing and experimentation, allowing researchers and developers to explore 5G functionalities without relying on proprietary systems \cite{12,4,5,6,7,21}. Open5GS provides a 5G core (5GC) network  implementation, UERANSIM serves as a UE and gNB simulator, srsRAN offers an open-source RAN, and OAI provides a full-stack 5G network solution. 

Research faces a major challenge of accurately simulating real-world network behavior, particularly in the domains of data analytics and mobility modeling \cite{13}. Real-world 5G deployments generate large volumes of operational data, and the ability to extract actionable insights from such data is critical to optimizing performance and managing resources efficiently \cite{36}. However, such data remain behind closed doors in proprietary systems.

NWDAF plays a central role in supporting closed-loop control in 5G networks by enabling network functions to expose and consume analytics services. By subscribing to events from network functions, NWDAF facilitates continuous monitoring of UE behavior, session lifecycle, and network state, while also generating predictive insights that help inform  decision-making. This makes NWDAF a cornerstone of emerging zero-touch network and service management framework, enabling networks to autonomously adapt to changing conditions without human intervention \cite{30,37}. However, in current open-source 5G ecosystems, NWDAF implementations remain relatively limited, often constrained to passive metrics collection. For example, while Free5GC supports some event exposure interfaces according to 3GPP specifications, its implementation lacks a complete notification mechanism to inform subscribers of event occurrences in real time. Similarly, the access and session management functions (AMF and SMF) in Free5GC do not natively support event subscriptions \cite{21}.

In this work, we address this gap by implementing a fully integrated, real-time NWDAF framework for Free5GC. Our design enhances the control plane with dynamic event subscription, notification delivery, and machine learning–based analytics for UE mobility and session dynamics. Our key contributions are as follows:
\begin{itemize}
    \item We develop an end-to-end NWDAF integration for Free5GC, supporting real-time event subscriptions and notifications from both AMF and SMF, adhering to 3GPP standards;
    \item We implement a custom activity-based mobility model (ABMM) within UERANSIM to simulate realistic, context-aware UE behavior that generates diverse network events;
    \item We incorporate a machine learning–driven module into the NWDAF to predict handover destinations based on historical UE behavior, achieving 80.65\% prediction accuracy;
    \item We validate our system through a two-week deployment with multiple virtual UEs and gNodeBs, demonstrating the utility of NWDAF for registration tracking, handover analysis, and predictive insights;
    \item We release our codebase and dataset as open-source to support future research on NWDAF capabilities in open 5G testbeds\footnote{https://github.com/HenokDanielbfg/5g-testbed-conference}.
\end{itemize}

The remainder of this paper is organized as follows: Section
\ref{sec-3}  reviews existing open-source 5G projects and related works on NWDAF implementations and integrations. Section \ref{sec-4} presents implementation details, including system architecture, data collection and event subscription mechanisms. Section \ref{sec-5}  discusses data generation results and network performance evaluation. Finally, Section \ref{sec-6} concludes the paper and highlights future directions.

\section{Related Work}\label{sec-3}
 
The transition to SBA in 5G core networks and the virtualization of radio and core networks have driven extensive research. This section explores current core and RAN implementations, and NWDAF development and integrations.

\subsection{Core and RAN implementation}

Several projects have emerged in the area of open source 5G network implementations, each contributing uniquely to the development and testing of 5G technologies.
Open5GS is an open-source project designed for building and managing new radio/long-term evolution (NR/LTE) mobile networks. It offers a comprehensive 5GC implementation, adhering to 3GPP Release 17 standards. Open5GS provides flexibility and scalability, making it suitable for both research and commercial deployments \cite{6,11}. 
UERANSIM  stands out as an open-source 5G UE and RAN simulator, offering simulation capabilities for both UE and gNodeB \cite{7}.
SrsRAN is an open-source project offering a complete suite for 4G and 5G RAN and core networks. It includes implementations for eNodeB (4G), gNodeB (5G), and UE, along with a core network for LTE \cite{4}. 
OpenAirInterface (OAI) provides open-source implementations for both 5GC and RAN. The OAI 5GC project aims to deliver a 3GPP-compliant 5G standalone core network with a rich feature set, while the RAN project focuses on implementing the 5G NR interface \cite{5}. 
Free5GC is an open-source implementation of the 5GC network based on 3GPP Release 16. It supports key functions such as the access and mobility management function (AMF), session management function (SMF), user plane function (UPF), and network repository function (NRF), and adopts a modular, service-based architecture (SBA) \cite{21}. Designed for research and prototyping, Free5GC enables RESTful inter-function communication.
IEEE 5G/6G Innovation Testbed is a cloud-based, end-to-end 5G network emulator developed by IEEE to facilitate the testing and experimentation of 5G and emerging 6G technologies \cite{3}. However, although it is composed of open-source components, the project itself is not freely available.
\subsection{NWDAF Integration}
A federated learning–based architecture for NWDAF is proposed in \cite{15} to mitigate overloading and data privacy issues. Their design features two components: leaf NWDAFs, deployed within each network function to create local models, and a root NWDAF, deployed in the core, which aggregates the locals into a global model. Notably, their work is conceptual, offering an architectural framework and use-case design.

Reference \cite{2} employ an NWDAF in analyzing a dataset from a simulated 5G network. The analysis particularly focuses on leveraging machine learning models to classify three kinds of network protocols. They tested 2 models (Random Forest and Decision Tree), and achieved an average accuracy of approximately 65\%.

In \cite{10}, the integration of NWDAF with Open5GS and UERANSIM is detailed. This work describes the collection of control-plane signaling data over a 138-minute period by leveraging the 5G service-based architecture to monitor inter-network function communications—such as BSF–NRF interactions—and using these insights to propose optimal placements for core network functions. The resulting dataset is open-sourced and available on GitHub.

In \cite{1}, the authors implement an NWDAF compatible with Free5GC. The NWDAF’s architecture aligns with that of the other Free5GC network functions. Their module comprises a model training logical function (MTLF) for training machine learning models and an analytics logical function (AnLF) for delivering inference results to consumers \cite{22}. They also successfully registered the NWDAF with the NRF. They utilize the MNIST dataset for analytics, leaving the integration with other network functions for live data collection and analytics as a future step. Their code is available on GitHub, which we have used as foundation for our current implementation.

This paper distinguishes itself by implementing a complete real-time event subscription notification system for both AMF and SMF, and an integrated NWDAF that collects data in real time and offers predictive analytics. While the referenced projects and NWDAF implementations offer valuable groundwork, they either lack end-to-end integration, real-time analytics capabilities, or open-source implementation. To our knowledge, this is the first open-source implementation of NWDAF in Free5GC that supports full end-to-end event monitoring, storage, and machine learning-driven analytics.
\section{Implementation}\label{sec-4}

\begin{figure}[]
    \centering
    \includegraphics[width=1\linewidth]{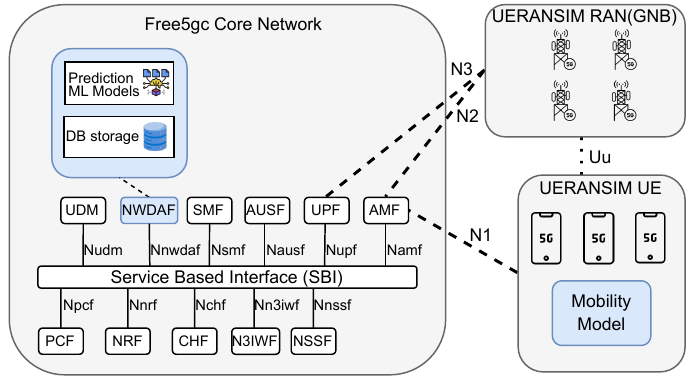}
    \caption{Network architecture}
    \label{fig:1}
\end{figure}

Our implementation integrates three main components: the 5GC network, RAN, and UE simulation, as shown in Fig. \ref{fig:1}. Each component is implemented using open-source software and configured to create a comprehensive testing environment. 
 
\subsection{Core Network Implementation}
The core network is implemented using Free5GC, an open source standalone (SA) 5G core network platform that adheres to 3GPP Release 15 specifications \cite{14,22}. This platform supports both experimental research and real-world deployments by offering a comprehensive suite of network functions, each responsible for specific tasks within the 5GC \cite{26}. In this architecture, the control plane operations, service discovery, and user plane traffic are efficiently managed through dedicated Network functions \cite{21}. 

We extend the existing Free5GC with event subscription mechanisms for the AMF and SMF. These functions expose RESTful APIs that allow other network functions to subscribe to specific network events. The AMF supports subscriptions for UE registration, deregistration, handover updates, and connectivity state changes. Similarly, the SMF provides subscription capabilities for session management events, including protocol data unit (PDU) establishment and modifications. 

\subsection{NWDAF}
\begin{figure}[]
    \centering
    \includegraphics[width=1\linewidth]{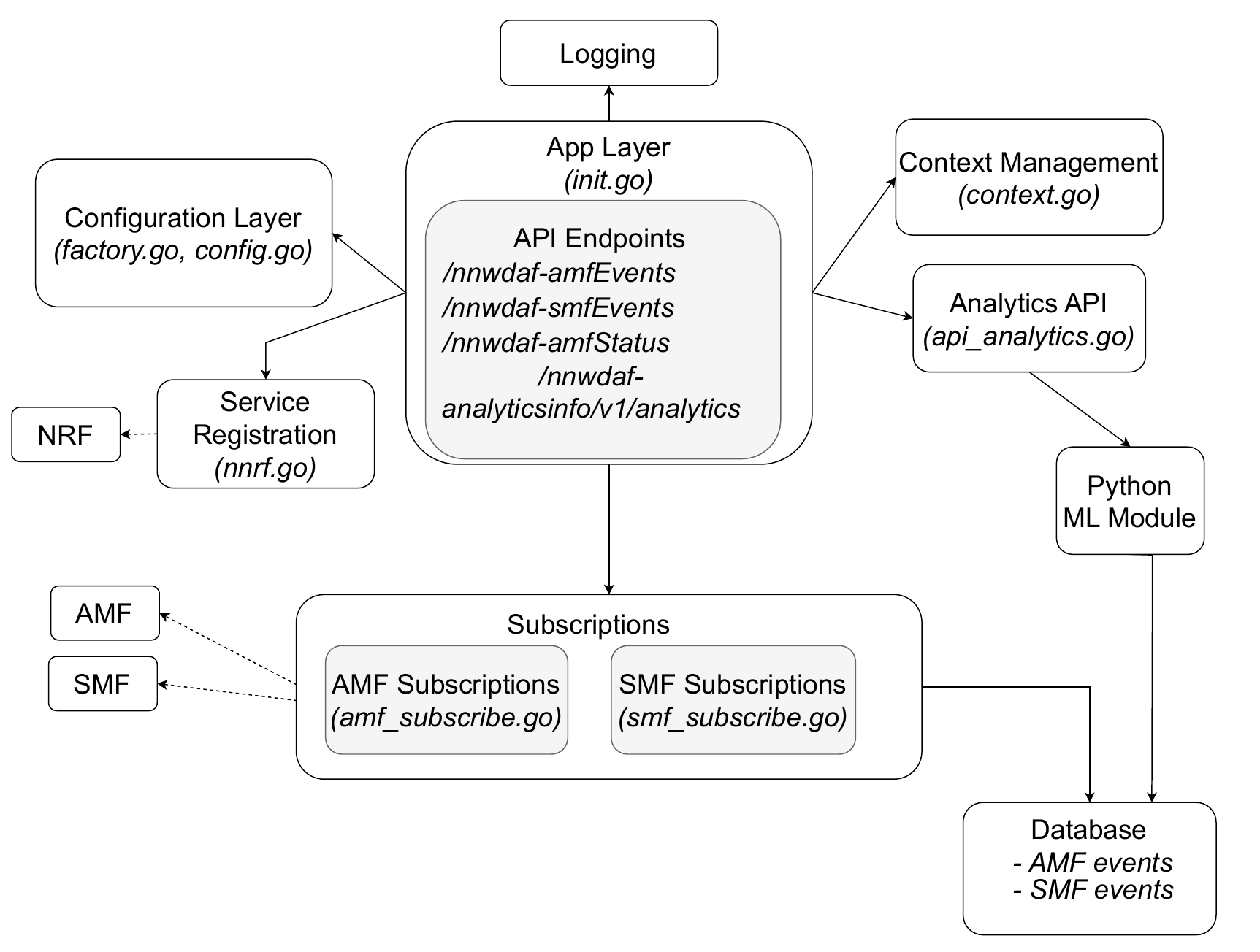}
    \caption{Proposed NWDAF architecture}
    \label{fig:2}
\end{figure}

\begin{figure}[]
    \centering
    \includegraphics[width=1\linewidth]{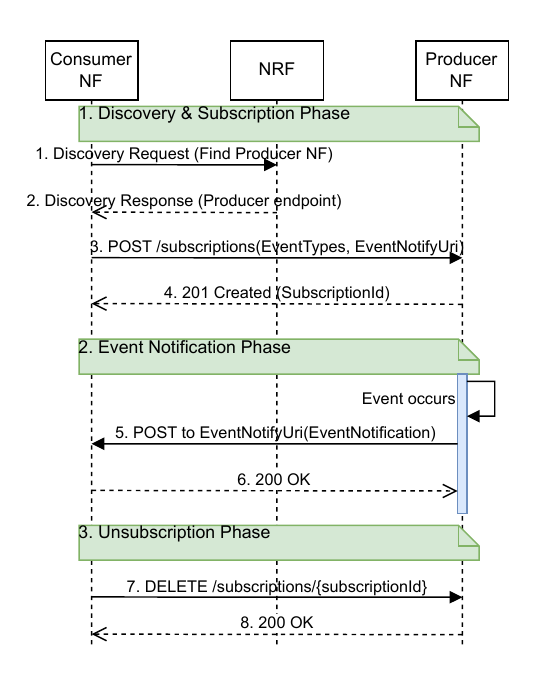}
    \caption{Subscription mechanism}
    \label{fig:3}
\end{figure}

The NWDAF architecture is shown in Fig.~\ref{fig:2}. The process of event subscriptions that the NWDAF utilizes is shown in Fig.~\ref{fig:3}.
Our implementation of the NWDAF builds on the work of Kim et al. who laid out the Free5GC based design of the network function \cite{1}. Our contribution is to include the addition of the “service” module, data collection mechanism, and prediction models. The NWDAF integrates with the core by first registering itself with the network repository function (NRF), where it advertises its services~\cite{29}. NWDAF reads from a configurable YAML file containing a list of events to subscribe to, pertaining to network functions of interest, and subscribes to said events via the consumer module. Our implementation currently supports subscriptions to AMF and SMF events.

\subsubsection{AMF Events Subscription}
In the Free5GC-based testbed, NFs query the NRF to discover the AMF’s event exposure service endpoint before subscribing to events via the "Namf\_EventExposure API", following the SBA.
Although Free5GC supports event subscription via the AMF, it lacks a built-in mechanism for notifying subscribers when events occur. To address this limitation, we implement a custom notification handler within the AMF that delivers event reports to registered consumers in real time.
Upon subscribing, the NWDAF provides a list of event types of interest, such as UE registration state, location reports, area presence, connectivity state, and UE reachability. The AMF logs each subscription, assigns a unique ID, and uses it for lifecycle management.
When a relevant event is triggered, the AMF invokes a custom "SendEventNotification()" function. This handler matches the event against active subscriptions and dispatches a structured report to each subscriber’s specified notification URI. During shutdown, NWDAF unsubscribes by issuing a delete request with the stored subscription ID to prevent orphaned entries.
\subsubsection{SMF Events Subscription}
Free5GC lacks default support for SMF event subscriptions. We extend its capabilities by implementing a complete subscription and notification mechanism via the "Nsmf\_EventExposure" API. NWDAF subscribes by sending a request specifying the desired event types:
PDU Session Establishment, PDU Session Release, Traffic Volume Reports, QoS Changes, UP Path Changes.
Upon receiving a request, SMF registers the subscriber and returns a unique subscription ID. Events, such as session creation or release, trigger "SendEventNotification()" function, that checks for matching subscriptions and sends a Notification to the subscriber’s provided URI. NWDAF handles these notifications for real-time analytics, enabling detection of session trends and network load conditions. To ensure clean shutdown, NWDAF unsubscribes by issuing a delete request, which invokes "SubscriptionsSubIdDelete()" to remove the record from SMF.
In addition to its event subscription and notification mechanisms, the NWDAF enhances the core network by integrating predictive analytics that forecast critical network events. By continuously analyzing both historical and recent data collected from network functions, the NWDAF has the ability to predict key parameters such as the most probable handover cell for user equipment, the expected duration of registration states, and future variations in active user counts. This forecasting layer empowers the network to proactively adjust resource allocation and optimize mobility management, thereby enhancing overall service quality.

\subsection{RAN \& UE Implementation}
The RAN is implemented using UERANSIM. It enables the deployment of multiple gNodeB instances, each configured with basic network parameters, such as frequency bands and cell configuration, and some capacity specifications, including connection handling capabilities, to simulate 5G protocol interactions. UERANSIM abstracts the physical layer while focusing on control plane and user plane procedures to test 5G network functionality \cite{7}. The system supports complete integration with the next generation core (NGC) network by handling control plane signaling via the NG application protocol (NGAP) over the N2 interface and managing user plane traffic through the GPRS tunneling protocol-user plane (GTP-U) protocol over the N3 interface, ensuring that the simulated network adheres closely to real-world 5G architectures\cite{24}. gNodeB instances are further designed to execute radio resource management, dynamically allocating resources based on quality of service (QoS) requirements and facilitating inter-gNodeB handovers to maintain continuous connectivity as UEs move across cells \cite{7}. 
In parallel, UERANSIM’s UE simulation capabilities enable the emulation of realistic user devices that support multiple PDU session types, including IPv4, IPv6, or dual-stack configurations, with each simulated UE operating with a complete 5G NAS protocol stack to accurately test processes such as network attachment, session establishment, and mobility management.

\subsubsection{Mobility Model}

To simulate realistic UE movement, we implement a custom activity-based mobility model within UERANSIM. This model captures human-like behavior by alternating between movement and dwell phases based on time-of-day and contextual preferences.

Each UE selects its next destination from a predefined set of activity locations, denoted as:
\[
\mathcal{L} = \{L_i\} = \left\{(x_i, y_i),\, \mu_i,\, \sigma_i \;\middle|\; i = 1,\dots,N \right\},
\]
where each location \(L_i\) is characterized by its coordinates \((x_i, y_i)\), and a dwell-time distribution modeled as \(\mathcal{N}(\mu_i, \sigma_i^2)\). Personal locations such as ``home'' and ``work'' are specified per the respective UE in its configuration file, while public destinations are shared, such as parks or coffee shops.
The day is divided into five time categories:
\[
t \in \{ \text{morning}, \text{lunch}, \text{afternoon}, \text{evening}, \text{night} \},
\]
Each category defines a probability distribution over activity types through a weight vector \(\mathbf{w}(t) = [w_1(t), w_2(t), \dots, w_K(t)]\). During morning hours, work-related destinations have higher weights, while recreational spots gain prominence in the evening.
When a UE is ready to select its next location:

\begin{enumerate}
    \item \textit{Activity Type Sampling:} A base type \(k\) is sampled according to \(\mathbf{w}(t)\), with a penalty applied to the current activity type \(k_0\) to reduce back-and-forth oscillations:
    \[
    w'_k(t) =
    \begin{cases}
    \epsilon \cdot w_k(t), & \text{if } k = k_0 \\
    w_k(t), & \text{otherwise}
    \end{cases}
    \]
    The adjusted weights are normalized to compute selection probabilities:
    \[
    P(k)=\frac{w'_k(t)}{\sum_j w'_j(t)}.
    \]
    
    \item \textit{Location and Movement:} A destination \(L_i\) is selected among those matching the sampled activity type. A base travel speed \(v_0\) is drawn from:
    \[
    v_0 \sim \mathcal{U}(v_{\min}, v_{\max}), \quad v = v_0 \cdot s(k, t),
    \]
    where \(s(k, t)\) adjusts speed to reflect contextual factors like location type and congestion.
    
    \item \textit{Navigation and Dwell:} The UE navigates toward the selected location using the direction:
    \[
    \theta = \tan^{-1} \left( \frac{y_{\text{dest}} - y_{\text{current}}}{x_{\text{dest}} - x_{\text{current}}} \right),
    \]
    and moves at speed \(v\). Upon arrival, the UE remains stationary for a duration $\tau \sim \mathcal{N}(\mu_i, \sigma_i^2)$.
\end{enumerate}

This model produces temporally grounded and spatially meaningful movement traces that reflect real-world human behavior. As reported by previous experimentations in the field, the resulting traces drive UE interactions such as registration, deregistration, and handover events \cite{ernesto1}, which are logged and analyzed by our NWDAF implementation.

\section{Data generation and Evaluation}\label{sec-5}

\begin{figure}[]
    \centering
    \includegraphics[width=0.85\linewidth]{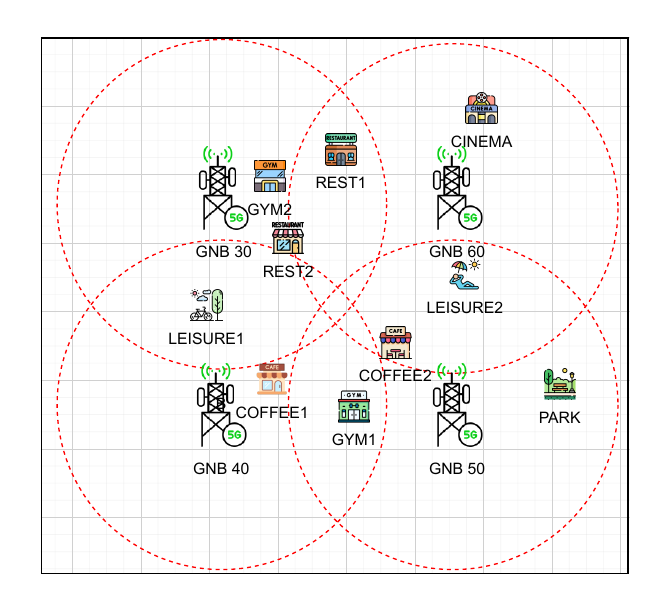}
    \caption{GNB and mobility location}
    \label{fig:5}
    
\end{figure}
We evaluate the NWDAF-enhanced 5G network testbed over a two-week period to analyze the behavior of the mobility model and assess the operational capabilities of the NWDAF. The testbed comprises four virtual gNodeBs, each assigned a distinct tracking area code (TAC) and arranged in a square formation, as illustrated in Fig.~\ref{fig:5}.
Each gNodeB is configured with a simulated coverage radius of 120 units, corresponding to a signal strength threshold of –120 dBm (consistent with UERANSIM’s default configuration). Four virtual UEs were deployed across the network: three configured to dynamically attach and detach from the network, and one configured to maintain continuous connectivity throughout the experiment. Upon joining, each UE begins from its designated home network. The simulation environment also includes several public locations, such as coffee shops and parks, that UEs can visit, following the defined activity-based mobility model.

\begin{figure}[]
    \centering
    \begin{subfigure}[b]{0.4\textwidth}
        \centering
        \includegraphics[width=\textwidth]{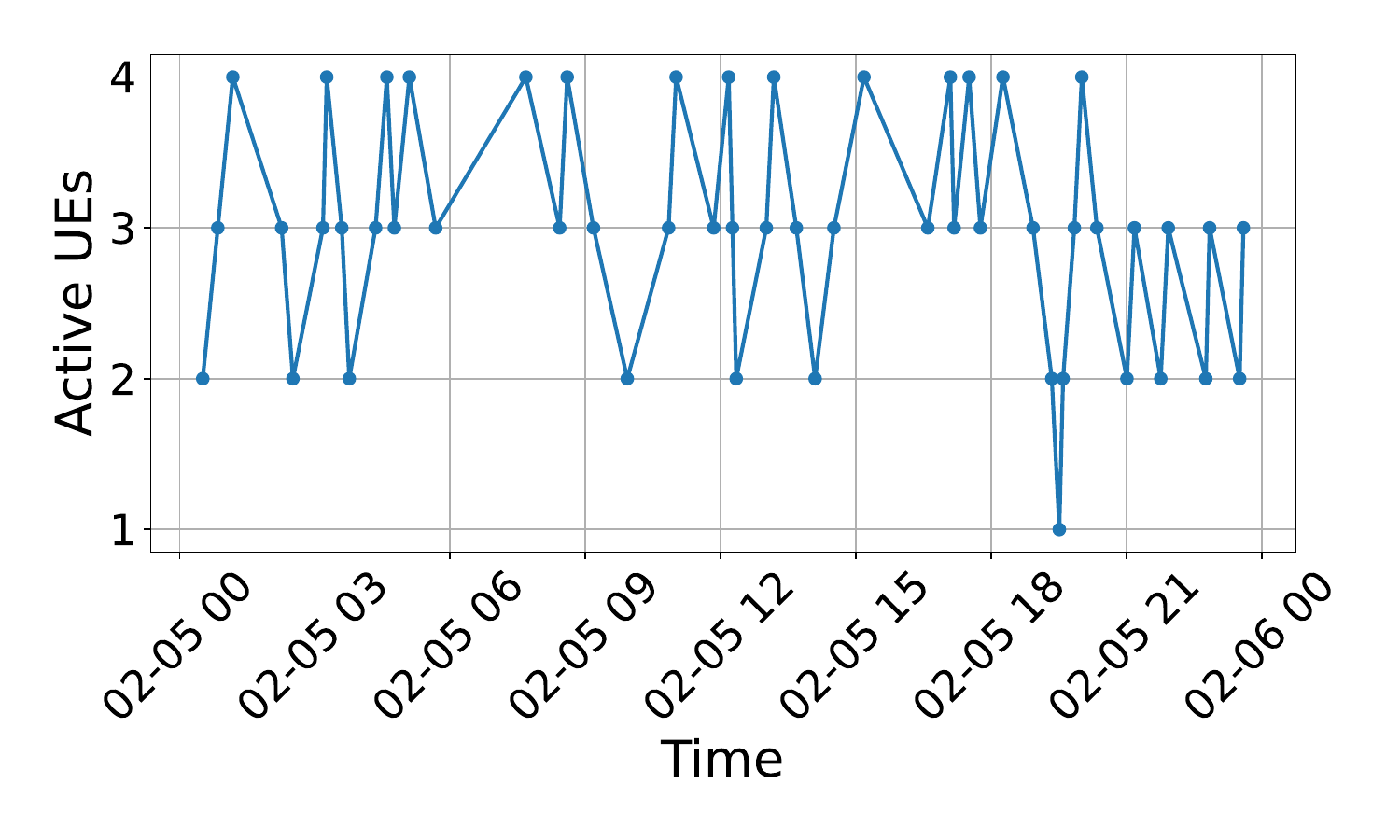}
        \caption{}
        \label{fig:active_ues}
    \end{subfigure}
    \hfill
    \begin{subfigure}[b]{0.45\textwidth}
        \centering
        \includegraphics[width=\textwidth]{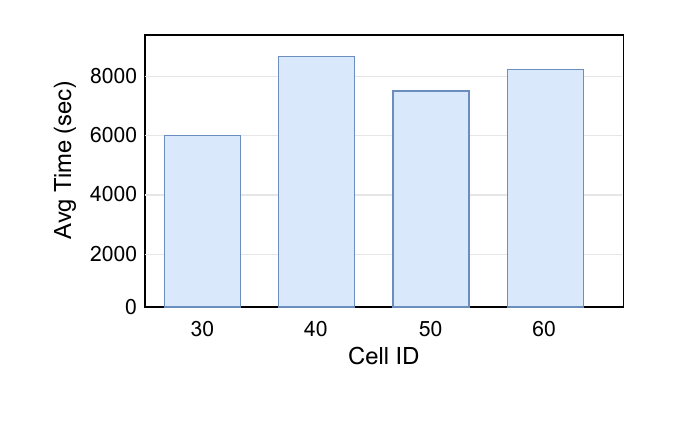}
        \caption{}
        \label{fig:avg_time}
    \end{subfigure}
    \hfill
    \begin{subfigure}[b]{0.45\textwidth}
        \centering
        \includegraphics[width=\textwidth]{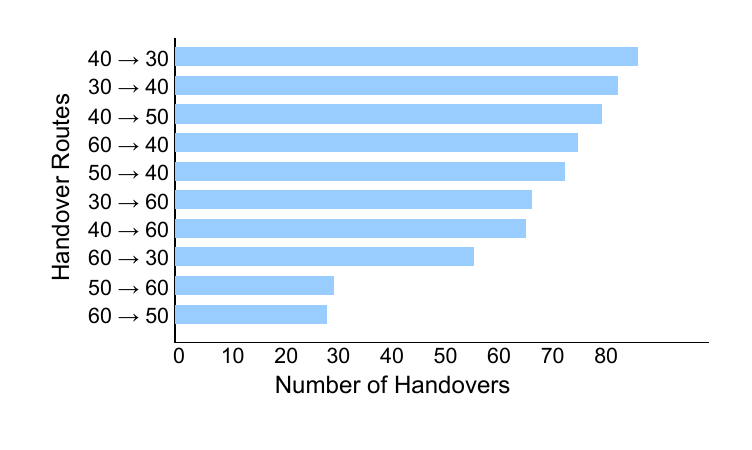}
        \caption{}
        \label{fig:handover_routes}
    \end{subfigure}
    \caption{Analysis of UE activity and handover behavior: (a) Number of active UEs throughout one day; (b) average time before handover; (c) handover routes}
    \label{fig:combined}
    \vspace{-2em}
\end{figure}
\subsection{UE Registration, Monitoring, and Handover Behavior}

Throughout the test period, comprehensive data were collected about UE registration, state transitions, and handovers to evaluate the network’s ability to analyze dynamic users behavior. All registration and deregistration attempts were successful, with the core network functions reliably executing mobility management procedures.
UEs exhibited distinct patterns in state durations, with the average time spent in the active state ranging from 100 to 102 minutes, while the inactive state typically lasted between 37 and 40 minutes.

Monitoring the number of active UEs over time showed clear temporal trends, which is expected due to the activity-based mobility model. The number of connected UEs varied throughout the day, with periods of high activity observed in the late morning around 11:00 and again between 14:00 and 17:00. While some hours had no active UEs, others reflected moderate load. The longest continuous activity session recorded for any UE was 9000 seconds (2.5 hours), demonstrating the system's ability to maintain stable connections over long durations.
Each handover event was logged with a timestamp, cell and device identifiers, and the TAC. These records indicate successful transitions across all configured cells in the testbed. One gNodeB emerged as a dominant traffic point, with significantly higher usage, due to its position with respect to the landmarks in the simulated topology. Handover frequency varied between UEs, with one device experiencing the most transitions, indicating either more aggressive movement or sensitivity to handover thresholds. Hourly trends shows that handovers are most common during mid-day and early evening hours, while early mornings remain relatively quiet. Certain cell pairs frequently exchanged UEs in both directions, pointing to either overlapping coverage zones or overly sensitive handover criteria. Dwell times within each cell also varied, with one particular cell showing the longest average residence time.
The NWDAF was actively engaged to collect analytics on UE mobility, registration events, and state transitions. Event subscription requests issued by the NWDAF were processed rapidly by the core network, with AMF and SMF acknowledging successful subscriptions in approximately 10ms. Moreover, The NWDAF received and processed event notifications in approximately 109 ms after the events occured. Performance monitoring revealed that the NWDAF operated with exceptional efficiency, utilizing merely 0.06\% of CPU resources and 0.17\% of system memory (27MB RAM). All generated insights were logged and stored, and the resulting dataset is made available for future use. While our current setup simulates a controlled mobility pattern, we note that more realistic models can be incorporated by others to stress-test the framework further. This dataset and framework can serve as a foundation for extending NWDAF research and exploring richer scenarios.
\subsection{Handover prediction model evaluation}
To evaluate the predictability of the UE handover events as a result of the activity-based mobility model, we employ four classification models on the UE location feature in the collected dataset, namely Random Forest, Gradient Boosting, $K$-Nearest Neighbors, and Decision Tree—on the dataset (UE location report). The input features include the UE’s subscriber ID (SUPI), the two most recently visited cells ID, time-of-day (the mobility model's time catgories), the (x, y) coordinates of the most recent (current) cell, and the frequency with which the UE visited that cell at the given time period.

We split the dataset into training and testing sets corresponding to 70\% and 30\%, respectively. Gradient Boosting with n\_estimators=100, max\_depth=9, and learning\_rate=0.05 produced the highest accuracy (80.65\%), followed by Random Forest (80.24\%), while Decision Tree and K-Nearest Neighbors were at 80.11\% and 79.03\%, respectively. These results suggest that the collected data exhibits predictable patterns, making it a valuable starting point for further NWDAF-driven exploration and analytics. The models we compared are known to be robust against  unreliability or low quality of training data \cite{ernesto2}.

\section{Conclusion}\label{sec-6}
We have presented an open-source, end-to-end implementation of the NWDAF integrated with Free5GC. Our framework extends the capabilities of existing open-source 5G core by introducing full support for AMF and SMF event subscriptions and real-time notification delivery, adhering to 3GPP’s service-based architecture. In addition, we have developed a lightweight, activity-based mobility model to simulate realistic UE behavior and trigger dynamic events across the network.
Through a two-week deployment involving multiple virtual gNodeBs and UEs, we demonstrated the effectiveness of our NWDAF in collecting, analyzing, and predicting mobility-related events. Both the NWDAF implementation and the resulting dataset have been open-sourced to support reproducibility and further exploration by the research community.

Future research can focus on extending NWDAF's capabilities to support additional event categories. Additional directions include implementing applications atop the NWDAF, such as traffic prediction and anomaly detection. Another promising direction involves integrating intent-driven query interfaces powered by natural language processing models, enabling operators and researchers to interact with NWDAF systems using intuitive, human-readable commands.
Moreover, Given the modular design and standards-compliant interfaces of our NWDAF framework, integrating it into large-scale experimentation platforms such as SLICES-RI would be an interesting path to explore. SLICES provides a pan-European infrastructure for reproducible and controlled experimentation in future networks, making it an ideal environment to deploy and evaluate our NWDAF in more diverse and geographically distributed 5G scenarios\cite{38}.

\bibliographystyle{IEEEtran}
\bibliography{references} 
\end{document}